\documentclass[11pt,a4paper]{article}

\usepackage[margin=2.5cm]{geometry}
\usepackage{amsmath,amssymb}
\usepackage{bm}
\usepackage{graphicx}
\usepackage{mathtools}
\usepackage{microtype}
\usepackage{xcolor}
\usepackage[numbers,sort&compress]{natbib}
\usepackage{authblk}
\usepackage[title]{appendix}
\usepackage[hidelinks]{hyperref}

\hypersetup{
  pdftitle={Extended-regularized lattice Boltzmann method},
  pdfauthor={Luís Orlando Emerich dos Santos, Diogo Nardelli Siebert}
}

\graphicspath{{figures/}}

\newcommand{\ee}{\mathrm{eq}}
\newcommand{\neqp}{\mathrm{neq}}
\newcommand{\nh}{\mathrm{nh}}
\newcommand{\Ma}{\mathrm{Ma}}
\newcommand{\Rey}{\mathrm{Re}}
\newcommand{\Order}{\mathcal{O}}
\newcommand{\Pc}{\mathsf{P}}
\newcommand{\Qc}{\mathsf{Q}}

\begin{document}

\title{Extended-regularized lattice Boltzmann method}

\author[1]{Lu\'is Orlando Emerich dos Santos\thanks{Corresponding author: \href{mailto:luis.emerich@ufsc.br}{luis.emerich@ufsc.br}}}
\author[1]{Diogo Nardelli Siebert\thanks{\href{mailto:diogo.siebert@ufsc.br}{diogo.siebert@ufsc.br}}}
\affil[1]{Department of Mobility Engineering, Federal University of Santa Catarina, Joinville, Santa Catarina, 89219-600, Brazil}
\date{}

\maketitle

\begin{abstract}
Regularized lattice Boltzmann methods improve numerical stability by projecting the nonequilibrium distribution onto its hydrodynamic second-order Hermite sector and filtering the remaining kinetic modes. We introduce a separate relaxation time for the complete complementary kinetic subspace. The resulting extended-regularized lattice Boltzmann method (E-RLBM) interpolates continuously between the Bhatnagar--Gross--Krook and regularized collision operators while retaining a two-rate implementation. A Chapman--Enskog expansion shows that the isothermal Navier--Stokes equations are recovered with kinematic viscosity $\nu=c_s^2(\tau-\Delta t/2)$, independently of the kinetic relaxation time $\tau_{\nh}$, whereas $\tau_{\nh}$ changes higher-order truncation errors. For linear transverse shear, its first effect is a fourth-derivative term whose coefficient depends on the direction of propagation relative to the lattice. This fourth-order coefficient is affine in $\tau_{\nh}$. For linear shear on D2Q9 and $\tau_{\nh}$ between $\tau$ and $\Delta t$, it lies between the corresponding BGK and RLBM values in every propagation direction. Stability maps for the decaying Taylor--Green vortex and flow past a rectangular obstacle show that an appropriate kinetic relaxation substantially enlarges the stable parameter domain. In a Taylor--Green convergence test with consistent initialization, all three operators converge at second order, and their error differences follow the hyperviscous coefficient for diagonal shear waves: the E-RLBM error is about $6\%$ above the BGK error and about $40\%$ below the RLBM error; in a double shear layer at $\Rey=80000$, the E-RLBM follows the RLBM convergence curve. The proposed collision operator therefore provides a low-cost means of controlling nonhydrodynamic modes without introducing a full multiple-relaxation-time basis.
\end{abstract}


\section{Introduction}
\label{sec:introduction}

The lattice Boltzmann method (LBM) represents fluid motion through the propagation and local collision of discrete particle distributions. Its kinetic formulation, geometrical flexibility, and locality make it attractive for simulations on parallel architectures \cite{McNamara1988,Higuera1989,Qian1992,Luo1997,Philippi2006,Shan2006,Krueger2017}. The widely used Bhatnagar--Gross--Krook (BGK) collision operator \cite{Bhatnagar1954,Qian1992} assigns a single relaxation time to all nonconserved moments. Although computationally economical, this coupling of hydrodynamic and kinetic modes can lead to poor stability at low viscosities and in under-resolved flows \cite{Lallemand2000,Luo2011,Coreixas2019,Coreixas2020,Wissocq2020}.

Several collision models have been proposed to address this limitation. Multiple-relaxation-time (MRT) methods separately control different moments \cite{Lallemand2000,dHumieres2002}. The two-relaxation-time (TRT) model provides a second rate without a full MRT basis, but separates even and odd velocity populations rather than the stress sector and its complete complement \cite{dHumieresGinzburg2009}. Central-moment and cumulant formulations change the representation in which relaxation is performed \cite{Geier2015}, and entropic schemes constrain the collision by kinetic entropy considerations \cite{Karlin1999,Ansumali2000}. Regularized lattice Boltzmann methods (RLBM) instead reconstruct the nonequilibrium distribution from the stress tensor and discard components outside the retained Hermite sector \cite{Ladd2001,Latt2006,Malaspinas2015,Coreixas2017,Mattila2017,Jonnalagadda2021}. This filtering can substantially improve stability, but the original regularized scheme fixes the relaxation of the discarded kinetic modes rather than treating it as an adjustable modeling parameter.

Here we introduce the extended-regularized lattice Boltzmann method (E-RLBM), which retains the full complementary nonequilibrium contribution and assigns it an independent relaxation time. The construction is a reduced MRT operator with one rate for the stress sector and one for all remaining kinetic modes. A related two-relaxation-time regularized formulation was recently introduced by Yu \textit{et al.} \cite{Yu2025}. Their scheme reconstructs selected higher-order nonequilibrium Hermite tensors and uses the additional rate to control boundary slip. The present formulation instead acts directly on the entire complement of the second-order projection; it therefore requires neither an explicit higher-order reconstruction nor a complete moment basis.

The study is organized around two questions suggested by the preliminary numerical results reported in Ref.~\cite{Santos2025}: Can the kinetic relaxation time be adjusted to enlarge the stable domain, and does this adjustment compromise hydrodynamic accuracy? Compared with the conference contribution, the present work provides the Chapman--Enskog analysis, identifies a $\tau_{\nh}$-dependent third-order contribution and the fourth-order correction for linear transverse shear, and adds a direct parameter sweep in a bounded flow. Section~\ref{sec:model} defines the collision operator and summarizes its macroscopic limit. Section~\ref{sec:numerical} describes the benchmarks, Sec.~\ref{sec:results} presents stability and accuracy results, and Appendix~\ref{app:ce} gives the Chapman--Enskog derivation.

\section{Kinetic-mode extended regularization}
\label{sec:model}

\subsection{General isothermal lattice and equilibrium}

Let $\{\bm{c}_i,w_i\}_{i=0}^{q-1}$ be an isothermal discrete-velocity quadrature in $d$ spatial dimensions, with lattice sound speed $c_s$. No particular velocity stencil or spatial dimension is assumed in the formulation below. The only lattice requirements are the standard quadrature symmetries needed by the second-order Hermite approximation. The distribution $f_i(\bm{x},t)$ evolves according to
\begin{equation}
f_i(\bm{x}+\bm{c}_i\Delta t,t+\Delta t)
=f_i(\bm{x},t)+\Omega_i(\bm{x},t),
\label{eq:lbe}
\end{equation}
where $\Omega_i$ is the collision operator. The density and momentum are
\begin{equation}
\rho=\sum_i f_i,
\qquad
\rho\bm{u}=\sum_i f_i\bm{c}_i.
\label{eq:hydro_fields}
\end{equation}

The second-order isothermal equilibrium is
\begin{equation}
f_i^{\ee}=w_i\rho
\left[
1+\frac{\bm{c}_i\cdot\bm{u}}{c_s^2}
+\frac{(\bm{c}_i\cdot\bm{u})^2}{2c_s^4}
-\frac{u^2}{2c_s^2}
\right].
\label{eq:equilibrium}
\end{equation}
Equations~\eqref{eq:lbe}--\eqref{eq:equilibrium}, as well as the projection and collision construction that follows, apply without modification to D2Q9, D3Q19, and D3Q27. These standard lattices have $c_s^2=1/3$ in lattice units; the derivation itself is expressed in terms of the general parameters $d$, $q$, $w_i$, and $c_s$.

\subsection{Collision operator}

Let $f_i^{\neqp}=f_i-f_i^{\ee}$ and let $f_i^*=f_i+\Omega_i$ denote the postcollision population. Following the operational notation used in the conference contribution \cite{Santos2025}, we split the nonequilibrium population into the stress-carrying regularized contribution $f_i^{(1)}$ and its complementary nonhydrodynamic contribution $f_i^{(>1)}$. The collision step is then written as
\begin{equation}
f_i^*=f_i^{\ee}
+\left(1-\frac{\Delta t}{\tau}\right)f_i^{(1)}
+\left(1-\frac{\Delta t}{\tau_{\nh}}\right)f_i^{(>1)},
\label{eq:postcollision}
\end{equation}
where
\begin{equation}
f_i^{(>1)}=f_i-f_i^{\ee}-f_i^{(1)}
=f_i^{\neqp}-f_i^{(1)}.
\label{eq:complement_conference}
\end{equation}
The regularized contribution is reconstructed from the nonequilibrium stress tensor as
\begin{equation}
f_i^{(1)}
=\frac{w_i}{2c_s^4}
\left(c_{i\alpha}c_{i\beta}-c_s^2\delta_{\alpha\beta}\right)
\Pi_{\alpha\beta}^{\neqp},
\label{eq:projection}
\end{equation}
where
\begin{equation}
\Pi_{\alpha\beta}^{\neqp}
=\sum_i f_i^{\neqp}c_{i\alpha}c_{i\beta}.
\end{equation}
Repeated Cartesian indices are summed. The superscript $(1)$ is the conventional regularization notation for the stress-carrying reconstruction; Chapman--Enskog orders are distinguished by square brackets in Appendix~\ref{app:ce}.

The same decomposition can be expressed compactly by defining the projectors
\begin{equation}
(\Pc f^{\neqp})_i\equiv f_i^{(1)},
\qquad
(\Qc f^{\neqp})_i\equiv f_i^{(>1)},
\qquad
\Qc=I-\Pc.
\label{eq:projectors}
\end{equation}
Because $f_i^{\neqp}$ has vanishing zeroth- and first-order moments, $\Qc f^{\neqp}$ contains no mass, momentum, or second-order stress contribution. Substituting Eq.~\eqref{eq:projectors} into Eq.~\eqref{eq:postcollision} gives the collision increment
\begin{equation}
\Omega_i
=-\frac{\Delta t}{\tau}(\Pc f^{\neqp})_i
-\frac{\Delta t}{\tau_{\nh}}(\Qc f^{\neqp})_i.
\label{eq:collision}
\end{equation}
The choices $\tau_{\nh}=\tau$ and $\tau_{\nh}=\Delta t$ recover, respectively, BGK and the original regularized operator. Collision-level damping of the kinetic sector requires
\begin{equation}
\left|1-\frac{\Delta t}{\tau_{\nh}}\right|<1,
\end{equation}
and hence $\tau_{\nh}>\Delta t/2$. This necessary local condition does not by itself guarantee nonlinear stability of the complete collision--streaming dynamics.

In $d$ dimensions, the conserved sector has $d+1$ modes, the symmetric second-order sector has $d(d+1)/2$ modes relaxed with $\tau$, and the remaining $q-(d+1)-d(d+1)/2$ kinetic modes are relaxed with $\tau_{\nh}$. Unlike a general MRT implementation, no explicit transformation to a prescribed moment basis is required. Because the trace of the nonequilibrium stress belongs to the second-order sector, its relaxation is tied to $\tau$; the present operator does not provide independent control of bulk viscosity.

Equivalently, Eq.~\eqref{eq:collision} is a reduced multiple-relaxation-time operator. The conserved, second-order, and complementary subspaces are mutually orthogonal with respect to the weighted inner product $\langle f,g\rangle=\sum_i f_ig_i/w_i$, and the nonconserved part relaxes with only two distinct rates: $\Delta t/\tau$ on the second-order Hermite subspace and $\Delta t/\tau_{\nh}$ on its complement. On the tensor-product lattices D2Q9 and D3Q27, the populations admit an orthogonal basis of independent discrete Hermite polynomials, formed by products of the one-dimensional polynomials $1$, $c$, and $c^2-c_s^2$; the complement is spanned by the members of order higher than two, and each of these moments relaxes with $\tau_{\nh}$ \cite{ShanChen2007}. On D3Q19, higher-order Hermite polynomials are not, in general, orthogonal to the second-order ones; for example, $\sum_iw_iH_{xxyy,i}H_{zz,i}=-1/27$, with $H_{xxyy}=(c_x^2-c_s^2)(c_y^2-c_s^2)$ and $H_{zz}=c_z^2-c_s^2$. The complement is then spanned by the higher-order polynomials after projection onto it, and an unmodified higher-order Hermite moment can contain a contribution of the second-order sector, which relaxes with $\tau$. It is not equivalent to the D2Q9 MRT model of Refs.~\cite{Lallemand2000,dHumieres2002} with the energy, energy-square, and heat-flux rates set to $s_e=\Delta t/\tau$ and $s_\varepsilon=s_q=\Delta t/\tau_{\nh}$. That model uses a moment basis that is orthogonal with respect to the unweighted inner product, and in that basis the trace mode of the second-order Hermite sector, $w_i(c_i^2-2c_s^2)$, has components of equal magnitude and opposite sign along the energy moment $e$ and the energy-square moment $\varepsilon$. The two operators therefore relax different combinations of the populations. The E-RLBM inherits the moment structure of regularization rather than that of the standard MRT, and it is defined without reference to a lattice-specific moment basis.

\subsection{Macroscopic limit and higher-order error}
\label{sec:macro_summary}

Appendix~\ref{app:ce} shows that Eq.~\eqref{eq:collision} recovers the isothermal weakly compressible Navier--Stokes equations,
\begin{equation}
\partial_t\rho+\partial_\alpha(\rho u_\alpha)=\Order(\epsilon^3),
\label{eq:continuity_summary}
\end{equation}
\begin{equation}
\partial_t(\rho u_\alpha)
+\partial_\beta(\rho u_\alpha u_\beta)
=-\partial_\alpha p
+\partial_\beta\left[
\mu(\partial_\alpha u_\beta+\partial_\beta u_\alpha)
\right]
+E_\alpha^{\mathrm{GI}}
+\Order(\epsilon^3),
\label{eq:momentum_summary}
\end{equation}
where $p=\rho c_s^2$ and
\begin{equation}
\nu=\frac{\mu}{\rho}
=c_s^2\left(\tau-\frac{\Delta t}{2}\right).
\label{eq:viscosity}
\end{equation}
The term $E_\alpha^{\mathrm{GI}}$ is the familiar cubic-velocity defect associated with the second-order equilibrium and is given explicitly in Eq.~\eqref{eq:gi_error}. It is of relative order $\Ma^2$ in the low-Mach regime.

The kinetic time $\tau_{\nh}$ does not enter Eq.~\eqref{eq:viscosity}. Appendix~\ref{app:ce} identifies a third-order contribution that vanishes for a linear transverse shear perturbation about rest. An independent long-wavelength expansion of that shear mode gives the bulk modified equation in lattice units,
\begin{equation}
\partial_t u_y
=\nu\partial_x^2u_y+\kappa_4\partial_x^4u_y
+\Order(\partial_x^6),
\label{eq:shear_modified_summary}
\end{equation}
with
\begin{equation}
\kappa_4
=\frac{a}{9}\left[a(2b-a)-\frac14\right],
\quad
a=\tau-\frac12,
\quad
b=\tau_{\nh}-\frac12.
\label{eq:kappa_summary}
\end{equation}
Equation~\eqref{eq:kappa_summary} holds for a wave vector aligned with a lattice axis. Appendix~\ref{app:ce} repeats the expansion on D2Q9 for a wave vector along a lattice diagonal, which gives
\begin{equation}
\kappa_4\!\left(\tfrac{\pi}{4}\right)=\frac{a^2}{18}\left(b-2a\right).
\label{eq:kappa_diag}
\end{equation}
Because the fourth-order term of the dispersion relation is a quartic form in the wave-vector components that must respect the square symmetry of the lattice, on D2Q9 a shear wave propagating at an angle $\theta$ to the $x$ axis obeys Eq.~\eqref{eq:shear_modified_summary}, with $u_y$ replaced by the transverse velocity component $u_\eta$, $\partial_x$ by the derivative along the propagation direction, and
\begin{equation}
\begin{aligned}
\kappa_4(\theta)&=\bar\kappa_4+\Delta\kappa_4\cos4\theta,\\
\bar\kappa_4&=\frac{a}{36}\left(5ab-4a^2-\frac12\right),\qquad
\Delta\kappa_4=\frac{a}{72}\left(6ab-1\right).
\end{aligned}
\label{eq:kappa_theta}
\end{equation}
The term $-a/36$ of Eq.~\eqref{eq:kappa_summary} is therefore absent along the diagonals, where $\kappa_4$ is only of order $a^2$, and the sign of the coefficient can change with direction: for $b>2a$, as in RLBM with $\tau<3/4$, the diagonal coefficient is positive. The fourth-order shear error is isotropic only when $b=1/(6a)$, which for BGK occurs at $\tau=1/2+1/\sqrt6$. For each direction, the fourth-derivative term is dissipative when $\kappa_4<0$. Its long-wavelength coefficient does not by itself determine the stability of modes near the grid scale or of a nonlinear bounded flow.

These fourth-order coefficients describe a one-parameter family that contains the two standard operators. At fixed $\tau$, Eq.~\eqref{eq:kappa_theta} is affine in $b$,
\begin{equation}
\kappa_4(\theta)=\kappa_4^{\mathrm{RLBM}}(\theta)+\left(b-\frac12\right)\frac{\partial\kappa_4}{\partial b},
\qquad
\frac{\partial\kappa_4}{\partial b}=\frac{a^2}{36}\left(5+3\cos4\theta\right)>0,
\label{eq:kappa_linear}
\end{equation}
and BGK and RLBM are the points $b=a$ and $b=1/2$ of the same straight line. The viscosity and $E^{\mathrm{GI}}_\alpha$ are common to the three operators. For this fourth-order linear shear coefficient on D2Q9, the E-RLBM reproduces the RLBM value for $\tau_{\nh}=\Delta t$ and the BGK value for $\tau_{\nh}=\tau$; for $\tau_{\nh}$ between these two values, its coefficient lies between those of BGK and RLBM in every propagation direction. Outside that interval it extrapolates them linearly, with a slope proportional to $a^2=9\nu^2$. Thus, within this long-wavelength linear shear analysis on D2Q9, the magnitude of the fourth-order coefficient cannot exceed the larger of the BGK and RLBM magnitudes when $\tau_{\nh}$ lies between $\tau$ and $\Delta t$. This bound does not extend to the total error of nonlinear flows or to boundary effects. Since the slope is positive in every direction, $\tau_{\nh}<\Delta t$ also makes the $K^4$ term more dissipative than in RLBM.

\section{Numerical methodology}
\label{sec:numerical}

The stability maps of Sec.~\ref{sec:results} were computed with the D3Q19 lattice on a single lattice layer in the $z$ direction, with periodic conditions in $z$, so that both flows are two-dimensional; the accuracy tests use D2Q9. Neither choice restricts the collision operator or its Chapman--Enskog limit. Because the regularized projection on D3Q19 also involves the $zz$ component of the nonequilibrium stress, these quasi-two-dimensional D3Q19 runs are not identical to D2Q9 runs for RLBM and E-RLBM, although they are for BGK. Unless stated otherwise, $\Delta x=\Delta t=1$. Periodic boundaries are used in both directions for the Taylor--Green and double-shear-layer problems. The characteristic time is $T=L/U_0$, and the Reynolds and Mach numbers are
\begin{equation}
\Rey=\frac{U_0L}{\nu},
\qquad
\Ma=\frac{U_0}{c_s}.
\end{equation}

\subsection{Stability benchmarks}

The decaying Taylor--Green vortex \cite{Taylor1923,Taylor1937} is initialized as
\begin{equation}
\begin{aligned}
u_x(x,y,0)&=U_0\sin(kx)\cos(ky),\\
u_y(x,y,0)&=-U_0\cos(kx)\sin(ky),
\end{aligned}
\label{eq:tgv_initial}
\end{equation}
where $k=2\pi/L$. For the stability map the populations are initialized at equilibrium with uniform density. The map uses $L=16$ and covers $0.05\leq\Ma\leq1$ and $10^{-5}\leq\nu\leq10^{-2}$ on a $400\times400$ grid of parameter values. A run is classified as unstable if its kinetic energy, $E=\frac12\sum_{\mathbf x}|\mathbf u(\mathbf x)|^2$, exceeds twice its initial value at any of the checks, performed every 100 time steps, during $100\,T$. The maps therefore record survival over this horizon under the stated criterion, not stability for unbounded times.

To explore the dependence on $\tau_{\nh}$, we also consider flow past a rectangular obstacle in a $20\times12$ domain at $\Ma=0.1$. A Zou--He velocity condition is imposed at the inlet \cite{Zou1997}. At the outlet, the populations are copied from the adjacent upstream node and rescaled so that the density equals its initial value, which gives a zero-normal-gradient condition for the velocity. The upper and lower boundaries are periodic, and no slip at the obstacle is imposed by halfway bounce-back. The obstacle is a $2\times4$ block, two nodes long in the flow direction and four nodes high, centered vertically; six fluid columns, including the inlet column, lie upstream of it and twelve, including the outlet column, downstream. The fluid starts at rest with uniform density and equilibrium populations. The stability survey varies $\tau_{\nh}$ from $0.5$ to $1.1$ and $\nu$ from $10^{-5}$ to $10^{-0.7}$ independently, on a $400\times400$ grid of parameter values. Each run is advanced for $200\,n_y/U_0$ time steps, where $n_y=12$, and is classified as unstable if, at any time step, the magnitude of the density at any node exceeds ten times its initial value, $|\rho|>10\rho_0$, or the density is not finite.

\subsection{Accuracy benchmarks}

Accuracy is assessed with two periodic problems of different character: the Taylor--Green vortex, a smooth flow with an exact solution, and the double shear layer, a strongly nonlinear flow without one, which is compared with a fine-grid reference.

The velocity field of Eq.~\eqref{eq:tgv_initial} is an exact solution of the two-dimensional incompressible Navier--Stokes equations. The advective term is balanced by the pressure gradient, and the vortex array decays by viscous diffusion without changing shape \cite{Taylor1923},
\begin{equation}
\begin{aligned}
u_x(x,y,t)&=U_0\sin(kx)\cos(ky)\,e^{-2\nu k^2t},\\
u_y(x,y,t)&=-U_0\cos(kx)\sin(ky)\,e^{-2\nu k^2t},
\end{aligned}
\label{eq:tgv_solution}
\end{equation}
with pressure $p=p_0+\frac14\rho_0U_0^2\left(\cos2kx+\cos2ky\right)e^{-4\nu k^2t}$. We use $\Rey=50$ with diffusive scaling: $\tau=0.62$ ($\nu=0.04$) is held fixed and $U_0=2/L$, so that $U_0=1/16$ at $L=32$ and the Mach number decreases in proportion to $1/L$. The grids are $L=32$--$256$. The error is measured at $t=T$, when the amplitude has decayed to $e^{-8\pi^2/\Rey}\approx0.21$ of its initial value for every $L$, and the reference field in Eq.~\eqref{eq:l2} is Eq.~\eqref{eq:tgv_solution}. The initial density follows from the exact pressure at $t=0$ through $p=\rho c_s^2$,
\begin{equation}
\rho=\rho_0\left[1+\frac{U_0^2}{4c_s^2}\left(\cos2kx+\cos2ky\right)\right],
\label{eq:tgv_density}
\end{equation}
and the populations are initialized consistently with the initial velocity gradient \cite{Mei2006}, $f_i=f_i^{\ee}+f_i^{(1)}$, where $f_i^{(1)}$ is given by Eq.~\eqref{eq:projection} with the Navier--Stokes stress $\Pi_{\alpha\beta}^{\neqp}=-\rho c_s^2\tau(\partial_\alpha u_\beta+\partial_\beta u_\alpha)$. For the Taylor--Green field $\partial_yu_x+\partial_xu_y=0$, so only $\Pi_{xx}^{\neqp}=-\Pi_{yy}^{\neqp}$ is nonzero. If the populations are initialized at equilibrium, the missing initial stress leaves, after the kinetic transient, a relative amplitude error whose leading long-wavelength term is $\tau(1-\tau)K^2/3$ in lattice units, where $K^2=2k^2$. This error is the same for the three operators and, at $\tau=0.62$, is roughly an order of magnitude larger than the operator-dependent error discussed in Sec.~\ref{sec:results}; with equilibrium initialization the errors in Fig.~\ref{fig:tgv_convergence} would be two to about four and a half times larger and would differ from one another by at most $20\%$.

The double shear layer \cite{Minion1997} consists of two thin shear layers of opposite sign in an $L\times L$ periodic domain. A small transverse perturbation makes them roll up into vortices, and at high Reynolds number the thin braids that form between them are under-resolved on coarse grids, which makes this flow a demanding test of both stability and accuracy. The initial velocity is
\begin{equation}
u_x=
\begin{cases}
U_0\tanh\!\left[\lambda(y/L-1/4)\right],&y\leq L/2,\\
U_0\tanh\!\left[\lambda(3/4-y/L)\right],&y>L/2,
\end{cases}
\label{eq:dsl_ux}
\end{equation}
\begin{equation}
u_y=\varepsilon U_0\sin\!\left[2\pi(x/L+1/4)\right],
\label{eq:dsl_uy}
\end{equation}
with $\lambda=80$, which sets the layer thickness, $\varepsilon=0.05$, $U_0=1/32$, and $\Rey=80000$. The density is initially uniform and the populations are initialized at equilibrium. The grid size ranges from $L=64$ to $1024$. Because $U_0$ is fixed, $\Ma\approx0.054$ on all grids, while $\nu=U_0L/\Rey$ gives $\tau-1/2=3\nu$ between $7.5\times10^{-5}$ at $L=64$ and $1.2\times10^{-3}$ at $L=1024$. An RLBM simulation with $L=2048$ supplies the reference field, and the error is evaluated at $t=T$ on the nodes of each grid, which coincide with every $(2048/L)$th node of the reference grid; no interpolation is used.

For both tests, the relative velocity error is
\begin{equation}
L_2=
\left[
\frac{\sum_{\bm{x}}\|\bm{u}_{\mathrm{sim}}-\bm{u}_{\mathrm{ref}}\|^2}
{\sum_{\bm{x}}\|\bm{u}_{\mathrm{ref}}\|^2}
\right]^{1/2}.
\label{eq:l2}
\end{equation}

\section{Results and discussion}
\label{sec:results}

\subsection{Stability}

\begin{figure}[t]
\includegraphics[width=0.8\textwidth]{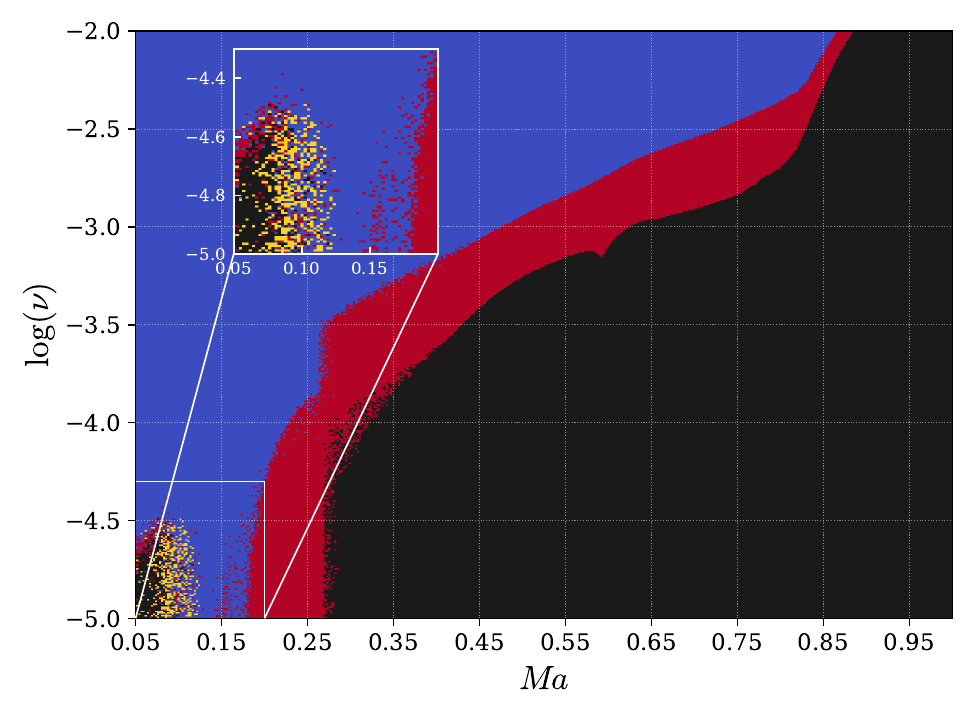}
\caption{Taylor--Green stability map for RLBM and E-RLBM with $\tau_{\nh}=0.65$. Horizontal axis: $\Ma$; vertical axis: $\log_{10}\nu$. Blue: both methods stable; red: only E-RLBM stable; black: both unstable; yellow: only RLBM stable. The inset enlarges the corner of low Mach number and low viscosity, which contains all 451 of the $160\,000$ parameter pairs that are stable only for RLBM.}
\label{fig:tgv_stability}
\end{figure}

Figure~\ref{fig:tgv_stability} compares the regularized method with the E-RLBM at $\tau_{\nh}=0.65$. The blue region contains cases for which both methods remain stable. The red band identifies cases stabilized only by the E-RLBM, while the black region is unstable for both. The adjustable kinetic relaxation extends the stability boundary across a broad range of Mach numbers, with the largest visible gain at intermediate viscosities. In the corner $\Ma\lesssim0.13$, $\log_{10}\nu\lesssim-4.4$, stable and unstable points of both methods are interspersed, and 451 points are stable only for RLBM. There, the instabilities grow slowly and reach the energy threshold close to the end of the $100\,T$ horizon, so the classification of individual points is sensitive to round-off; recompiling the code changes the classification of some of these points without altering the overall pattern.

\begin{figure}[tbp]
\includegraphics[width=0.8\textwidth]{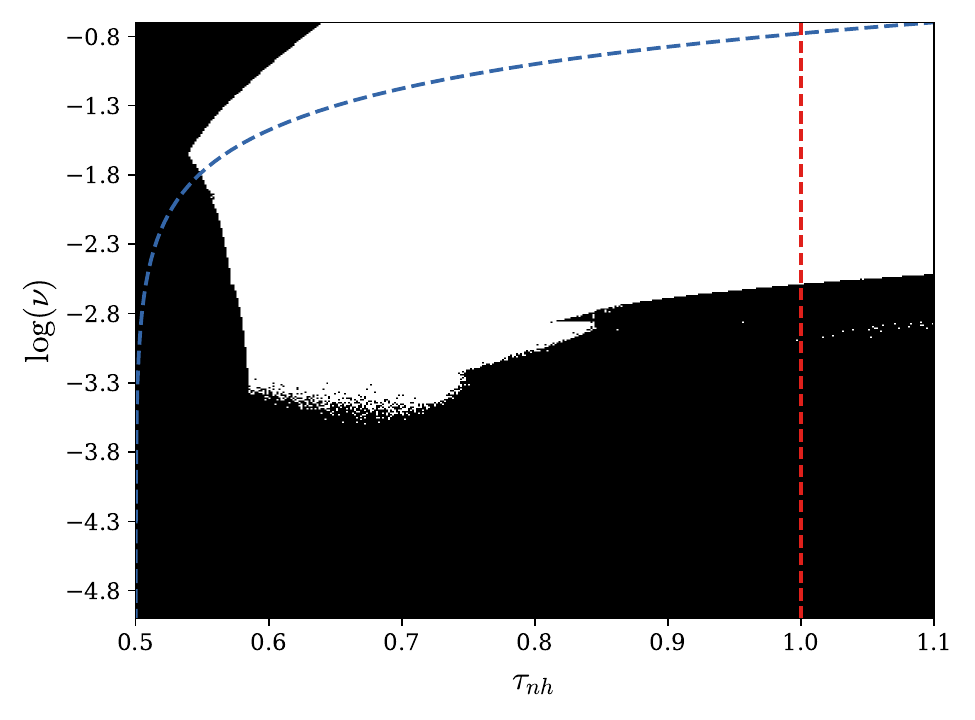}
\caption{Stability map for flow past a rectangular obstacle at $\Ma=0.1$. Horizontal axis: $\tau_{\nh}$; vertical axis: $\log_{10}\nu$. White and black denote stable and unstable simulations, respectively. The red line marks RLBM ($\tau_{\nh}=1$); the blue dashed curve marks BGK ($\tau_{\nh}=\tau$).}
\label{fig:obstacle_stability}
\end{figure}

The obstacle-flow sweep in Fig.~\ref{fig:obstacle_stability} makes the parameter dependence more explicit. The white region is stable and the black region unstable. The vertical red line marks the regularized value $\tau_{\nh}=1$, whereas the blue dashed curve is the BGK locus $\tau_{\nh}=\tau=1/2+3\nu$. The stable domain is largest for an over-relaxed kinetic sector, approximately $0.6\lesssim\tau_{\nh}\lesssim0.8$ over the low-viscosity part of the map. The result also shows that the best value is problem- and resolution-dependent; no single kinetic relaxation eliminates all stability limitations.

The modified equation establishes unchanged leading-order viscosity, but its long-wavelength hyperviscous term cannot quantitatively explain the stability maps. At fixed $\tau$, the change from RLBM ($\tau_{\nh}=1$) to $\tau_{\nh}=0.65$ is $\Delta\kappa_4=-0.7\nu^2$ along a lattice axis and $-0.175\nu^2$ along a diagonal, in lattice units. For $\nu=10^{-3}$ and $10^{-5}$, the axis change is only about $0.85\%$ and $0.0084\%$ of $|\kappa_4|$, respectively; along the diagonals, $\kappa_4$ is itself of order $\nu^2$ for both operators. Moreover, $\kappa_4$ varies monotonically with $\tau_{\nh}$, whereas Fig.~\ref{fig:obstacle_stability} has a bounded stability window. The gain must therefore involve finite-wavelength collision--streaming dynamics and, in the obstacle test, may also involve the boundary treatment. In particular, RLBM sets the complementary nonequilibrium sector to zero after collision, while $\tau_{\nh}=0.65$ multiplies it by approximately $-0.54$; stronger local damping of that sector cannot explain the gain. A full spectral and boundary analysis is beyond the scope of these maps.

\subsection{Accuracy and convergence}

\begin{figure}[tbp]
\includegraphics[width=0.8\textwidth]{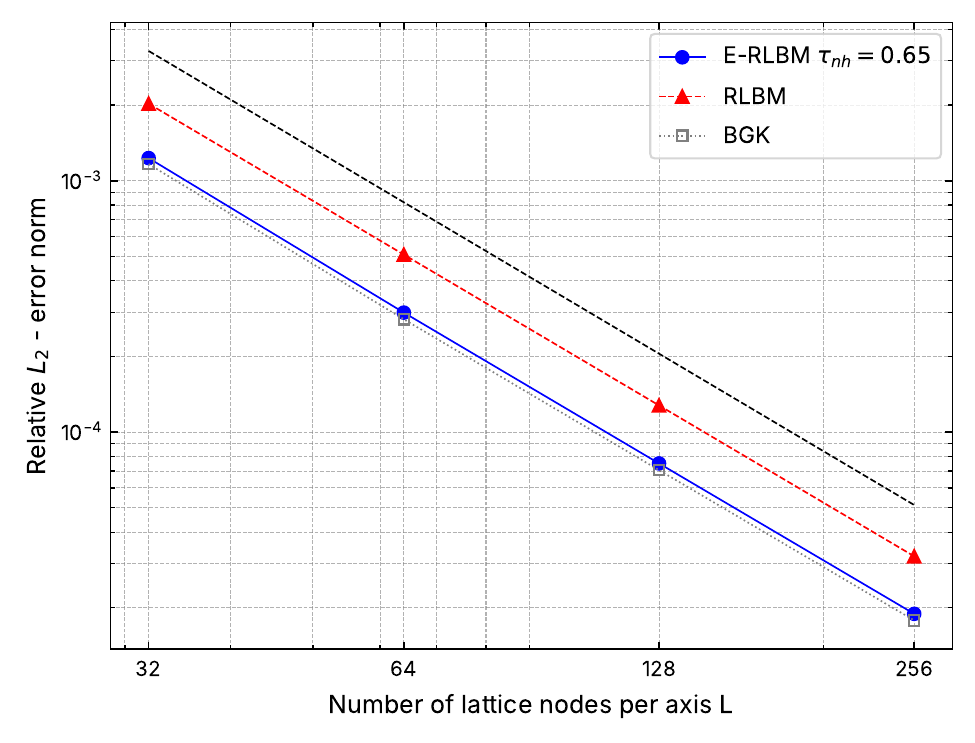}
\caption{Relative velocity error, Eq.~\eqref{eq:l2}, for the Taylor--Green vortex at $\Rey=50$ and $t=T$, with $\tau=0.62$ fixed, $U_0=2/L$, and populations initialized as $f_i^{\ee}+f_i^{(1)}$. The E-RLBM uses $\tau_{\nh}=0.65$. The black dashed line indicates an $L^{-2}$ slope.}
\label{fig:tgv_convergence}
\end{figure}

Figure~\ref{fig:tgv_convergence} compares BGK, RLBM, and E-RLBM for the Taylor--Green vortex at $\Rey=50$. All three operators converge at second order: the error decreases by a factor of four at each refinement. At every resolution the E-RLBM error is about $6\%$ above the BGK error, and the RLBM error is about $80\%$ above it. These differences are explained by the hyperviscous term. The Taylor--Green velocity field is a superposition of shear waves with wave vectors $(\pm k,\pm k)$, that is, along the lattice diagonals, so the relevant coefficient is Eq.~\eqref{eq:kappa_diag}. At $\tau=0.62$ it gives $\kappa_4(\pi/4)=-9.6\times10^{-5}$ for BGK, $-7.2\times10^{-5}$ for E-RLBM, and $+2.1\times10^{-4}$ for RLBM. In diffusive scaling the hyperviscous term changes the decay rate by the relative amount $-\kappa_4(\pi/4)K^2/\nu=a(2a-b)K^2/6$, an $L^{-2}$ contribution like the $\Order(\Ma^2)$ error, which is common to the three operators and slows the decay. The negative coefficients of BGK and E-RLBM partly cancel the $\Ma^2$ error, whereas the positive RLBM coefficient adds to it. At $L=256$, this estimate predicts error differences from BGK of $1.14\times10^{-6}$ for E-RLBM and $1.45\times10^{-5}$ for RLBM, against $1.11\times10^{-6}$ and $1.42\times10^{-5}$ in the simulations. We also checked Eq.~\eqref{eq:kappa_diag} in the linear regime ($U_0=10^{-3}$, $L=64$) by measuring the decay rate for $0.6\leq\tau\leq1.2$: the relative viscosity error of the three operators agrees with $a(2a-b)K^2/6$ within $4\%$, except near $\tau=3/4$ for RLBM, where the predicted value vanishes and the deviation is below $10^{-6}$. The E-RLBM coefficient differs from the BGK and RLBM coefficients by $a^2(b-a)/18$ and $a^2(b-1/2)/18$, respectively. With $\tau=0.62<\tau_{\nh}=0.65<1$ it lies between the two, close to BGK; the measured error also lies between the parent errors, consistent with the interpolation of the leading shear coefficient in Eq.~\eqref{eq:kappa_linear}. For diagonal linear shear with $a>0$ and $b>0$, the magnitude of the E-RLBM hyperviscous coefficient exceeds that of BGK when $b<a$ or $b>3a$; the magnitudes are equal at $b=a$ and $b=3a$. For $\tau\to1/2$, the differences among the three coefficients vanish as $a^2$. The kinetic relaxation therefore leaves the order of accuracy unchanged in this test and, for the chosen parameters, places the fourth-order shear coefficient between the values of the two limiting operators.

\begin{figure}[t]
\includegraphics[width=0.8\textwidth]{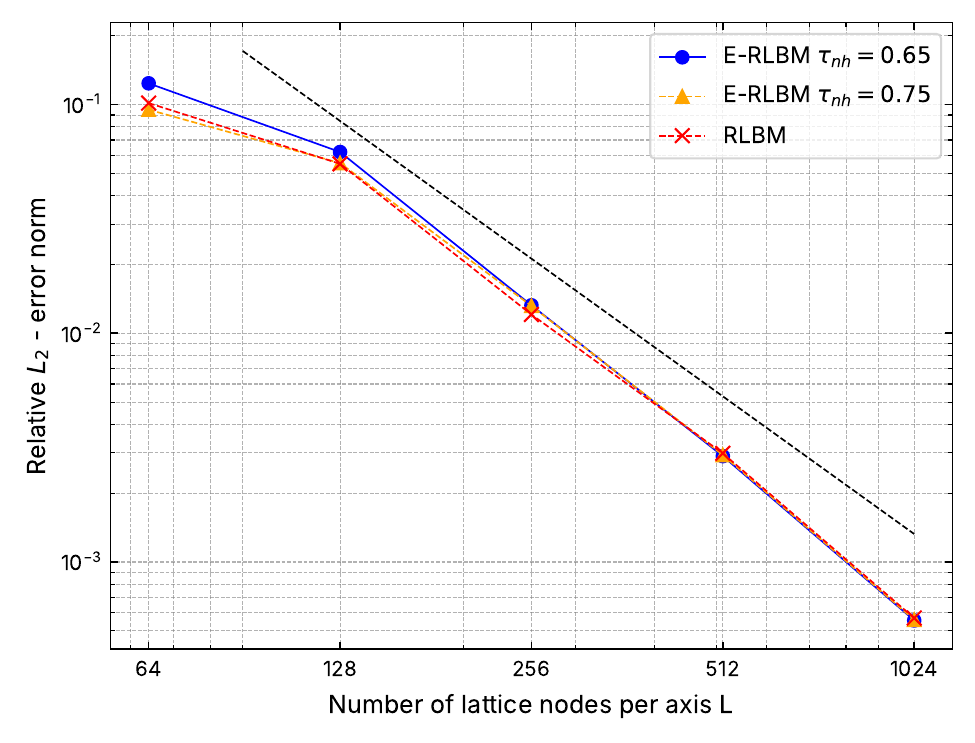}
\caption{Relative velocity error, Eq.~\eqref{eq:l2}, for the double shear layer at $\Rey=80000$, $U_0=1/32$, and $t=T$. The reference is an RLBM simulation with $L=2048$. The black dashed line indicates an $L^{-2}$ slope.}
\label{fig:dsl_convergence}
\end{figure}

\begin{figure*}[t]
\includegraphics[width=0.64\textwidth]{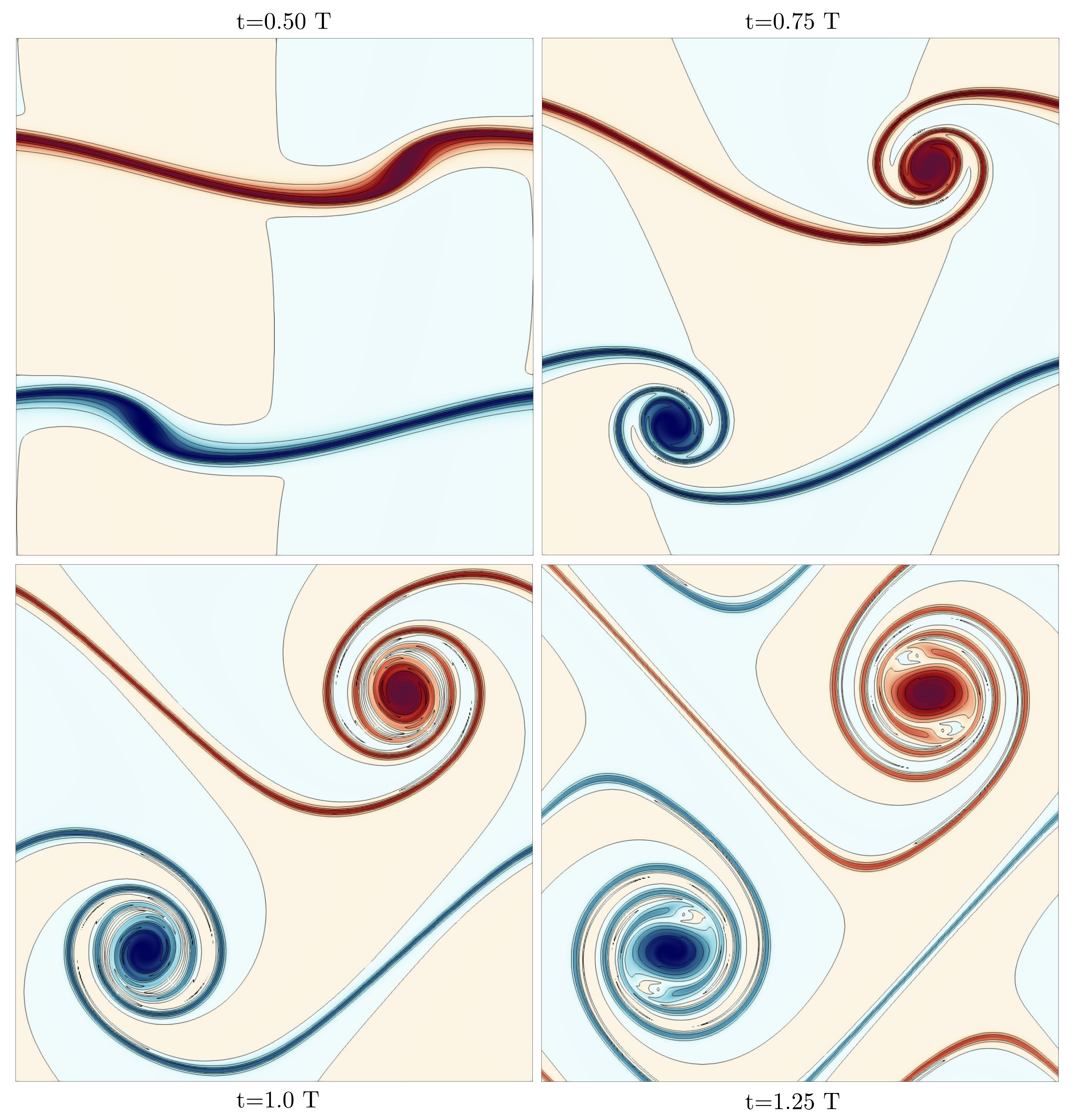}
\caption{Representative double-shear-layer evolution computed with the E-RLBM ($\tau_{\nh}=0.75$) at $L=512$, $U_0=1/32$, and $\Rey=80000$. The panels correspond to $t/T=0.50$, $0.75$, $1.00$, and $1.25$; color denotes signed vorticity.}
\label{fig:dsl_evolution}
\end{figure*}

The double-shear-layer problem provides a more demanding nonlinear test. The representative evolution in Fig.~\ref{fig:dsl_evolution} develops the expected roll-up of both shear layers. At $\Rey=80000$, Fig.~\ref{fig:dsl_convergence} shows that E-RLBM with $\tau_{\nh}=0.65$ and $0.75$ follows the RLBM convergence curve. On the coarsest grid, $\tau_{\nh}=0.65$ gives a larger error than RLBM, whereas $\tau_{\nh}=0.75$ gives a slightly smaller one; from $L=256$ onward the three errors differ by at most about $10\%$, and by less than $3\%$ at $L=512$ and $1024$. Between $L=128$ and $512$ the observed order of convergence is close to two. Between $L=512$ and $1024$ it rises to about $2.4$, which is consistent with a second-order error measured against a reference that is only twice as fine as the finest grid. Since this reference is produced by one of the compared methods, the test measures relative agreement within this family rather than an independent error against Navier--Stokes. For comparison, BGK becomes unstable at $L=64$ before $t=T$.

\section{Conclusions}
\label{sec:conclusions}

We introduced a two-rate extension of the regularized lattice Boltzmann collision operator. The nonequilibrium stress sector relaxes with $\tau$, while the complete complementary kinetic sector relaxes with $\tau_{\nh}$. This decomposition preserves the implementation simplicity of regularization and avoids the full moment-space machinery of a general MRT model.

The Chapman--Enskog analysis establishes that the Navier--Stokes viscosity depends only on $\tau$, whereas $\tau_{\nh}$ enters higher-order errors. For linear transverse shear, its first effect is the fourth-order hyperviscous coefficient, which depends on the propagation direction. The E-RLBM contains RLBM ($\tau_{\nh}=\Delta t$) and BGK ($\tau_{\nh}=\tau$) as particular cases. For linear shear on D2Q9, the hyperviscous coefficient is affine in $\tau_{\nh}$: for $\tau_{\nh}$ between $\tau$ and $\Delta t$ it lies between the BGK and RLBM values in every propagation direction, and it differs from either by an amount proportional to $\nu^2$. This bounds the magnitude of that fourth-order coefficient by the larger of the two parent magnitudes, but does not establish a general bound on the total error of nonlinear flows or on boundary effects. Taylor--Green and obstacle-flow stability maps show that selected values enlarge the stable region. The accuracy tests retain second-order convergence: in the Taylor--Green test, the E-RLBM error lies between those of BGK and RLBM, $6\%$ above the former and well below the latter, with differences that follow the diagonal hyperviscous coefficient. In the double shear layer, both tested kinetic times give errors within $3\%$ of RLBM at $L=512$ and $1024$, although $\tau_{\nh}=0.65$ produces larger errors on the coarser grids. For the tested parameters, these results indicate a stability gain at an accuracy comparable to that of the parent operators, without establishing a universal optimum or attributing the stability gain to the long-wavelength hyperviscous term.

The optimal kinetic relaxation is not universal. It depends on the lattice, resolution, flow, and boundary treatment. Future work should combine the present bulk analysis with a full-wavenumber spectral analysis, wall-slip measurements, comparisons with TRT and hybrid regularized models, and independent high-resolution references.

\section*{Acknowledgements}
The authors gratefully acknowledge the support of Equinor Brazil through Project ANP 23270-2 and the support of the Brazilian National Agency of Petroleum, Natural Gas and Biofuels (ANP) through the research and development levy.

\section*{Data availability}
The source code of all simulations reported in this paper, the SimBoltz library functions it uses, the stability maps, and the convergence data are available at \url{https://github.com/loesantos/ERLBM}. The full simulation fields, including those used in Fig.~\ref{fig:dsl_evolution}, are available from the corresponding author upon reasonable request.

\section*{Competing interests}
The authors declare that they have no known competing financial interests or personal relationships that could have appeared to influence the work reported in this paper.

\section*{Use of generative AI tools}
During the preparation of this work the authors used Claude (Anthropic) and ChatGPT (OpenAI) in order to improve the language of the manuscript, to assist in revising the presentation of the analysis, to check algebraic derivations with symbolic computation, and to prepare scripts for plotting the figures. After using these tools, the authors reviewed and edited the content as needed and take full responsibility for the content of this article.

\begin{appendices}
\numberwithin{equation}{section}
\section{Chapman--Enskog analysis}
\label{app:ce}

This appendix derives the bulk macroscopic equations associated with
Eq.~\eqref{eq:collision} and isolates a third-order truncation contribution that
depends on $\tau_{\nh}$. The transverse shear analysis below is a separate
fourth-order calculation. The analysis assumes smooth fields, no external force,
constant relaxation times $\tau$ and $\tau_{\nh}$, and an isothermal lattice satisfying the quadrature symmetries required by the
second-order equilibrium. Boundary-induced errors are not included. To
distinguish the conventional regularization symbol $f_i^{(1)}$ in
Eq.~\eqref{eq:projection} from the Chapman--Enskog coefficients, square
brackets label asymptotic orders.

\subsection{Collision projectors and multiscale expansion}

Let
\begin{equation}
g_i=f_i-f_i^{\ee},
\qquad
s_\nu=\frac{\Delta t}{\tau},
\qquad
s_{\nh}=\frac{\Delta t}{\tau_{\nh}}.
\end{equation}
Because $f_i$ and $f_i^{\ee}$ have the same conserved moments,
$\sum_i g_i=0$ and $\sum_i c_{i\alpha}g_i=0$. The projectors are used on this
nonconserved subspace, on which
\begin{equation}
\Pc^2=\Pc,
\qquad
\Qc^2=\Qc,
\qquad
\Pc\Qc=\Qc\Pc=0,
\qquad
\Pc+\Qc=I.
\end{equation}
The collision operator is
\begin{equation}
\Omega_i=-s_\nu(\Pc g)_i-s_{\nh}(\Qc g)_i.
\label{eq:collision_appendix}
\end{equation}
Both projected contributions have vanishing zeroth- and first-order moments,
whereas $\Qc g$ additionally has a vanishing second-order moment. It is useful
to introduce the relaxation-time operator
\begin{equation}
\mathsf{T}=\tau\Pc+\tau_{\nh}\Qc,
\qquad
\mathsf{T}^{-1}=\frac{1}{\tau}\Pc+\frac{1}{\tau_{\nh}}\Qc,
\label{eq:Toperator}
\end{equation}
where the inverse is understood on the nonconserved subspace.

We use the acoustic Chapman--Enskog scaling
\begin{align}
f_i&=f_i^{[0]}+\epsilon f_i^{[1]}+\epsilon^2f_i^{[2]}
+\epsilon^3f_i^{[3]}+\cdots,\label{eq:ce_f_expansion}\\
\partial_t&=\epsilon\partial_{t_1}+\epsilon^2\partial_{t_2}
+\epsilon^3\partial_{t_3}+\cdots,\label{eq:ce_t_expansion}\\
\partial_\alpha&=\epsilon\partial_{\alpha1},
\label{eq:ce_x_expansion}
\end{align}
and define
\begin{equation}
D_{1i}=\partial_{t_1}+c_{i\alpha}\partial_{\alpha1}.
\end{equation}
All conserved fields are assigned to $f_i^{[0]}$ by imposing
\begin{equation}
\sum_i f_i^{[n]}=0,
\qquad
\sum_i c_{i\alpha}f_i^{[n]}=0,
\qquad n\geq1.
\label{eq:ce_gauge}
\end{equation}

With $D_i=\partial_t+c_{i\alpha}\partial_\alpha$, Taylor expansion of
Eq.~\eqref{eq:lbe} gives
\begin{equation}
\Delta tD_if_i+\frac{\Delta t^2}{2}D_i^2f_i
+\frac{\Delta t^3}{6}D_i^3f_i+\cdots=\Omega_i.
\label{eq:taylor_lbe}
\end{equation}

\subsection{Zeroth and Euler orders}

At order $\epsilon^0$, the collision term requires
\begin{equation}
f_i^{[0]}=f_i^{\ee}.
\end{equation}
At order $\epsilon^1$, after division by $\Delta t$,
\begin{equation}
D_{1i}f_i^{\ee}
=-\frac{1}{\tau}(\Pc f^{[1]})_i
-\frac{1}{\tau_{\nh}}(\Qc f^{[1]})_i.
\label{eq:ce_order1}
\end{equation}
Before inverting the collision operator, its conserved moments provide the
solvability conditions. The zeroth and first moments of
Eq.~\eqref{eq:ce_order1} are
\begin{align}
\partial_{t_1}\rho+\partial_{\alpha1}(\rho u_\alpha)&=0,
\label{eq:euler_mass}\\
\partial_{t_1}(\rho u_\alpha)
+\partial_{\beta1}\Pi_{\alpha\beta}^{\ee}&=0,
\label{eq:euler_momentum}
\end{align}
where
\begin{equation}
\Pi_{\alpha\beta}^{\ee}
=\sum_i f_i^{\ee}c_{i\alpha}c_{i\beta}
=\rho u_\alpha u_\beta+p\delta_{\alpha\beta},
\qquad p=\rho c_s^2.
\label{eq:pi_eq_appendix}
\end{equation}
Once Eqs.~\eqref{eq:euler_mass} and \eqref{eq:euler_momentum} are satisfied,
$D_1f^{\ee}$ belongs to the nonconserved subspace and Eq.~\eqref{eq:ce_order1}
can be inverted:
\begin{equation}
f_i^{[1]}
=-\tau(\Pc D_1f^{\ee})_i
-\tau_{\nh}(\Qc D_1f^{\ee})_i.
\label{eq:f1}
\end{equation}

Define the equilibrium third-order moment and its first-order balance by
\begin{align}
Q_{\alpha\beta\gamma}^{\ee}
&=\sum_i f_i^{\ee}c_{i\alpha}c_{i\beta}c_{i\gamma},
\label{eq:qeq_definition}\\
A_{\alpha\beta}
&=\partial_{t_1}\Pi_{\alpha\beta}^{\ee}
+\partial_{\gamma1}Q_{\alpha\beta\gamma}^{\ee}.
\label{eq:Adef}
\end{align}
Since $\Qc f^{[1]}$ has no second-order moment, Eq.~\eqref{eq:f1} gives
\begin{equation}
\Pi_{\alpha\beta}^{[1]}
=\sum_i f_i^{[1]}c_{i\alpha}c_{i\beta}
=-\tau A_{\alpha\beta}.
\label{eq:pi1}
\end{equation}
Although $f_i^{[1]}$ depends on both relaxation times, its second-order moment
therefore depends only on $\tau$.

\subsection{Navier--Stokes order}

At order $\epsilon^2$, Eq.~\eqref{eq:taylor_lbe} becomes
\begin{equation}
\partial_{t_2}f_i^{\ee}+D_{1i}f_i^{[1]}
+\frac{\Delta t}{2}D_{1i}^2f_i^{\ee}
=-(\mathsf{T}^{-1}f^{[2]})_i.
\label{eq:ce_order2}
\end{equation}
The zeroth moment, together with the Euler solvability conditions and
Eq.~\eqref{eq:ce_gauge}, yields
\begin{equation}
\partial_{t_2}\rho=0.
\end{equation}
For the first moment, the identity
\begin{equation}
\sum_i c_{i\alpha}D_{1i}^2f_i^{\ee}
=\partial_{\beta1}A_{\alpha\beta}
\label{eq:D1sq_identity}
\end{equation}
follows by applying $D_{1i}$ once more to the Euler equations. Hence,
\begin{equation}
\partial_{t_2}(\rho u_\alpha)
+\partial_{\beta1}\Pi_{\alpha\beta}^{[1]}
+\frac{\Delta t}{2}\partial_{\beta1}A_{\alpha\beta}=0.
\end{equation}
Using Eq.~\eqref{eq:pi1}, this can be written as
\begin{equation}
\partial_{t_2}(\rho u_\alpha)
+\partial_{\beta1}\left[
\left(1-\frac{\Delta t}{2\tau}\right)
\Pi_{\alpha\beta}^{[1]}
\right]=0,
\label{eq:momentum_order2}
\end{equation}
or, equivalently,
\begin{equation}
\partial_{t_2}(\rho u_\alpha)
=\partial_{\beta1}\left[
\left(\tau-\frac{\Delta t}{2}\right)A_{\alpha\beta}
\right].
\label{eq:momentum_order2b}
\end{equation}
This proves that $\tau_{\nh}$ does not enter the Navier--Stokes transport
coefficient.

For the second-order equilibrium in Eq.~\eqref{eq:equilibrium}, the third-order
moment is
\begin{equation}
Q_{\alpha\beta\gamma}^{\ee}
=p\left(
u_\alpha\delta_{\beta\gamma}
+u_\beta\delta_{\alpha\gamma}
+u_\gamma\delta_{\alpha\beta}
\right).
\label{eq:qeq}
\end{equation}
In particular, it lacks the Maxwell--Boltzmann cubic contribution
$\rho u_\alpha u_\beta u_\gamma$. Using the Euler equations in
Eq.~\eqref{eq:Adef} gives
\begin{equation}
A_{\alpha\beta}
=p(\partial_{\alpha1}u_\beta+\partial_{\beta1}u_\alpha)
-\partial_{\gamma1}(\rho u_\alpha u_\beta u_\gamma).
\label{eq:Aeval}
\end{equation}
Combining the $t_1$ and $t_2$ scales recovers
Eqs.~\eqref{eq:continuity_summary}--\eqref{eq:momentum_summary}, with
\begin{equation}
\mu=p\left(\tau-\frac{\Delta t}{2}\right),
\qquad
\nu=\frac{\mu}{\rho}
=c_s^2\left(\tau-\frac{\Delta t}{2}\right),
\label{eq:viscosity_appendix}
\end{equation}
and the cubic-velocity defect
\begin{equation}
E_\alpha^{\mathrm{GI}}
=-\left(\tau-\frac{\Delta t}{2}\right)
\partial_\beta\partial_\gamma
(\rho u_\alpha u_\beta u_\gamma).
\label{eq:gi_error}
\end{equation}
Relative to the viscous term, Eq.~\eqref{eq:gi_error} scales as
$U^2/c_s^2=\Ma^2$.

For later use, inversion of Eq.~\eqref{eq:ce_order2} gives
\begin{equation}
f^{[2]}
=-\mathsf{T}\left[
\partial_{t_2}f^{\ee}+D_1f^{[1]}
+\frac{\Delta t}{2}D_1^2f^{\ee}
\right].
\label{eq:f2}
\end{equation}

\subsection{First kinetic-mode-dependent correction}

At order $\epsilon^3$, the complete distribution equation is
\begin{align}
\partial_{t_3}f^{\ee}
+\partial_{t_2}f^{[1]}
+D_1f^{[2]}
&+\frac{\Delta t}{2}\left(
D_1^2f^{[1]}+2D_1\partial_{t_2}f^{\ee}
\right)\notag\\
&+\frac{\Delta t^2}{6}D_1^3f^{\ee}
=-\mathsf{T}^{-1}f^{[3]}.
\label{eq:ce_order3}
\end{align}
We isolate the part generated by the complementary kinetic sector by defining
\begin{equation}
\chi_i=(\Qc D_1f^{\ee})_i,
\qquad
B_{\alpha\beta}
=\sum_i c_{i\alpha}c_{i\beta}D_{1i}\chi_i.
\label{eq:Bdef}
\end{equation}
After the Euler solvability conditions are applied, $\chi_i$ has vanishing
moments through second order. Therefore,
\begin{equation}
B_{\alpha\beta}
=\partial_{\gamma1}
\sum_i c_{i\alpha}c_{i\beta}c_{i\gamma}\chi_i,
\qquad
\sum_i c_{i\alpha}D_{1i}^2\chi_i
=\partial_{\beta1}B_{\alpha\beta}.
\label{eq:Bidentities}
\end{equation}

Three terms in the first moment of Eq.~\eqref{eq:ce_order3} contribute to the
coefficient of $\partial_{\beta1}B_{\alpha\beta}$. First, the part of the
second-order stress generated by $\chi_i$ is
\begin{equation}
\left.\Pi_{\alpha\beta}^{[2]}\right|_\chi
=\tau\left(\tau_{\nh}-\frac{\Delta t}{2}\right)
B_{\alpha\beta}.
\label{eq:pi2_chi}
\end{equation}
Second, since $(f^{[1]})_{\Qc}=-\tau_{\nh}\chi$, the corresponding part of the
$D_1^2f^{[1]}$ term is
\begin{equation}
\left.
\frac{\Delta t}{2}
\sum_i c_{i\alpha}D_{1i}^2f_i^{[1]}
\right|_\chi
=-\frac{\Delta t\,\tau_{\nh}}{2}
\partial_{\beta1}B_{\alpha\beta}.
\label{eq:D1sq_f1_chi}
\end{equation}
Finally, the complementary part of the cubic Taylor term is
\begin{equation}
\left.
\frac{\Delta t^2}{6}
\sum_i c_{i\alpha}D_{1i}^3f_i^{\ee}
\right|_\chi
=\frac{\Delta t^2}{6}
\partial_{\beta1}B_{\alpha\beta}.
\label{eq:D1cube_feq_chi}
\end{equation}
Their sum is
\begin{align}
\Theta
&=\tau\left(\tau_{\nh}-\frac{\Delta t}{2}\right)
-\frac{\Delta t\,\tau_{\nh}}{2}
+\frac{\Delta t^2}{6}\notag\\
&=\left(\tau-\frac{\Delta t}{2}\right)
\left(\tau_{\nh}-\frac{\Delta t}{2}\right)
-\frac{\Delta t^2}{12}.
\label{eq:theta}
\end{align}
Consequently, one bulk error contribution at order $\epsilon^3$ controlled by the kinetic
relaxation is
\begin{equation}
E_\alpha^{\nh}
=-\Theta\,\partial_{\beta1}B_{\alpha\beta}.
\label{eq:enh}
\end{equation}
Equation~\eqref{eq:enh} is not the complete $\epsilon^3$ equivalent equation:
terms independent of $\tau_{\nh}$, including the remaining Burnett-order
contributions, have not been displayed. Before lower-order time derivatives
are eliminated, Eq.~\eqref{eq:enh} contains third-order derivatives. This contribution vanishes for the linear transverse shear mode studied below;
it cannot be used to derive that mode's fourth-derivative coefficient.

\subsection{Transverse shear modified equation}
\label{app:shear}

The $\tau_{\nh}$-dependent part of the third-order equation, Eq.~\eqref{eq:enh},
does not act on a linear transverse shear perturbation about rest. In this
geometry $\partial_y=0$ and $u_x=0$. Since the third moment of
$\Pc D_1 f^{\ee}$ vanishes by parity, Eq.~\eqref{eq:Bdef} gives
$B_{yx}=\partial_{x1}\{(1-c_s^2)\partial_{x1}(\rho u_xu_y)\}$, which is quadratic
in the perturbation and vanishes because $u_x=0$. The first effect of
$\tau_{\nh}$ on this flow is therefore of fourth order in the gradients.
Carrying the multiscale expansion to $\epsilon^4$ would be lengthy. For a small
perturbation of a uniform fluid at rest, however, the lattice Boltzmann equation
is linear with constant coefficients, and its equivalent equation can be read
directly from the evolution of a single Fourier mode, in the spirit of the
linear analysis of Ref.~\cite{Lallemand2000}. The procedure has four steps:
(i) linearize collision and streaming about rest; (ii) identify a small set of
moments that is closed under both steps for a transverse wave; (iii) write one
time step of a Fourier mode as a $3\times3$ recurrence and expand its
hydrodynamic eigenvalue $\lambda(k)$ in powers of the wavenumber $k$; and
(iv) compare $\lambda(k)$ with the growth factor of a modified equation. The only
small parameter is $k\Delta x$; no separation of time scales and no ordering of
time derivatives is required. Lattice units, $\Delta x=\Delta t=1$ and
$c_s^2=1/3$, are used throughout this subsection, together with
\begin{equation}
a=\tau-\frac12,
\qquad
b=\tau_{\nh}-\frac12,
\qquad
\omega_1=1-\frac1\tau,
\qquad
\omega_2=1-\frac1{\tau_{\nh}}.
\end{equation}

\paragraph{Linearization}
Let $f_i=w_i\rho_0+\delta f_i$, where $\delta f_i$ is small, and retain only
terms linear in $\delta f_i$. The perturbations of density and velocity are
$\delta\rho=\sum_i\delta f_i$ and $\rho_0\delta u_\alpha=\sum_ic_{i\alpha}\delta f_i$,
and the linearized equilibrium is
$\delta f_i^{\ee}=w_i\left(\delta\rho+\rho_0c_{i\alpha}\delta u_\alpha/c_s^2\right)$.
Its second moment is $\delta\Pi^{\ee}_{\alpha\beta}=c_s^2\delta\rho\,\delta_{\alpha\beta}$, and its
third moment is
\begin{equation}
\sum_i\delta f_i^{\ee}c_{i\alpha}c_{i\beta}c_{i\gamma}
=\rho_0c_s^2\left(\delta u_\alpha\delta_{\beta\gamma}+\delta u_\beta\delta_{\alpha\gamma}+\delta u_\gamma\delta_{\alpha\beta}\right),
\label{eq:lin_q_eq}
\end{equation}
which requires only the fourth-order isotropy of the lattice. The collision,
Eq.~\eqref{eq:collision_appendix}, becomes
\begin{equation}
\delta f_i^*=\delta f_i^{\ee}+\omega_1\left(\Pc\,\delta f^{\neqp}\right)_i+\omega_2\left(\Qc\,\delta f^{\neqp}\right)_i,
\qquad
\delta f^{\neqp}=\delta f-\delta f^{\ee},
\label{eq:lin_collision}
\end{equation}
and streaming is $\delta f_i'(\mathbf x)=\delta f_i^*(\mathbf x-\mathbf c_i)$. For a
Fourier mode $\delta f_i\propto\exp(\mathrm{i}\mathbf k\cdot\mathbf x)$, streaming
multiplies the amplitude of population $i$ by $\exp(-\mathrm{i}\mathbf k\cdot\mathbf c_i)$.
One time step is thus a linear map $\hat f\mapsto M(\mathbf k)\hat f$, the product of
the $\mathbf k$-independent collision matrix and a diagonal matrix of phases. At
$\mathbf k=0$, the eigenvalues of $M$ are $1$ for the conserved moments, $\omega_1$
for the stress, and $\omega_2$ for the complementary sector. The hydrodynamic
eigenvalues are the branches with $\lambda\to1$ as $|\mathbf k|\to0$; the others
describe kinetic modes. At zero wavenumber, these modes decay when
$|\omega_1|,|\omega_2|<1$, although this decay can be arbitrarily slow as either
modulus approaches one.

Two properties of the projectors make the reduction below possible. By
Eq.~\eqref{eq:projection}, $\Pc\,\delta f^{\neqp}$ has the same second moments as
$\delta f^{\neqp}$, and, being proportional to $w_iH_{i\alpha\beta}$, it has no odd
moments. Hence $\Qc\,\delta f^{\neqp}$ has no second moments. Consequently, any
component of the stress relaxes with $\omega_1$, and any third-order moment
relaxes toward its equilibrium value with $\omega_2$:
\begin{equation}
\Pi_{\alpha\beta}^*=\Pi_{\alpha\beta}^{\ee}+\omega_1\left(\Pi_{\alpha\beta}-\Pi_{\alpha\beta}^{\ee}\right),
\qquad
Q_{\alpha\beta\gamma}^*=Q_{\alpha\beta\gamma}^{\ee}+\omega_2\left(Q_{\alpha\beta\gamma}-Q_{\alpha\beta\gamma}^{\ee}\right),
\label{eq:lin_moment_relax}
\end{equation}
where $\Pi$ and $Q$ denote the second and third moments of the perturbation.
Equation~\eqref{eq:lin_moment_relax} is the only place where the E-RLBM differs
from BGK ($\omega_2=\omega_1$) and RLBM ($\omega_2=0$) in the calculation.

\paragraph{Wave vector along a lattice axis}
Let $\mathbf k=(k,0,0)$ and group the populations by their $x$ velocity,
\begin{equation}
g_{c}=\sum_{i:\,c_{ix}=c}c_{iy}\,\delta f_i,\qquad c\in\{-1,0,1\}.
\end{equation}
The three moments
\begin{equation}
\begin{aligned}
m_0&=\sum_i c_{iy}\,\delta f_i=g_{-1}+g_0+g_1,\\
m_1&=\sum_i c_{ix}c_{iy}\,\delta f_i=g_1-g_{-1},\\
m_2&=\sum_i c_{ix}^2c_{iy}\,\delta f_i=g_1+g_{-1}
\end{aligned}
\label{eq:shear_moments}
\end{equation}
determine $(g_{-1},g_0,g_1)$ and conversely. Here $m_0=\rho_0\delta u_y$ is the
transverse momentum, $m_1=\Pi_{xy}$ is the shear stress, and $m_2=Q_{xxy}$ is
the third-order moment that carries the $x$ flux of that stress. From the
linearized equilibrium and Eq.~\eqref{eq:lin_q_eq},
\begin{equation}
m_1^{\ee}=0,
\qquad
m_2^{\ee}=\rho_0c_s^2\delta u_y=\frac{m_0}{3},
\end{equation}
on D2Q9, D3Q19, and D3Q27. Equation~\eqref{eq:lin_moment_relax} then gives the
collision step
\begin{align}
m_0^*&=m_0,\label{eq:shear_collision_m0}\\
m_1^*&=\omega_1m_1,
\label{eq:shear_collision_m1}\\
m_2^*&=\frac{m_0}{3}+\omega_2\left(m_2-\frac{m_0}{3}\right).
\label{eq:shear_collision_m2}
\end{align}
Streaming multiplies $g_c$ by $\exp(-\mathrm{i}kc)$. With $C_k=\cos k$ and
$S_k=\sin k$, and expressing the $g_c$ through Eq.~\eqref{eq:shear_moments},
\begin{align}
m_0'&=m_0^*+(C_k-1)m_2^*-\mathrm{i}S_km_1^*,
\label{eq:shear_stream_m0}\\
m_1'&=C_km_1^*-\mathrm{i}S_km_2^*,
\label{eq:shear_stream_m1}\\
m_2'&=C_km_2^*-\mathrm{i}S_km_1^*.
\label{eq:shear_stream_m2}
\end{align}
Equations~\eqref{eq:shear_collision_m0}--\eqref{eq:shear_stream_m2} form a closed
$3\times3$ recurrence: the post-collision moments depend only on
$(m_0,m_1,m_2)$, and the post-streaming moments only on their post-collision
values. No other moment feeds back into them, whatever the remaining
populations do. On D3Q19 and D3Q27 the group $c_{ix}=0$ contains several
populations, but they do not move in $x$ and enter only through their sum
$g_0$; the recurrence is therefore the same on the three lattices. Its terms
have a direct interpretation: $-\mathrm{i}S_km_1^*\approx-\partial_x\Pi_{xy}$ in
Eq.~\eqref{eq:shear_stream_m0} is the divergence of the stress;
$-\mathrm{i}S_km_2^*\approx-\partial_xQ_{xxy}$ in Eq.~\eqref{eq:shear_stream_m1}
produces stress from the gradient of the third moment, which at equilibrium is
$-c_s^2\rho_0\partial_x\delta u_y$; and Eq.~\eqref{eq:shear_stream_m2} feeds the
stress back into the third moment, which relaxes with $\tau_{\nh}$.

\paragraph{Hydrodynamic eigenvalue}
A mode of the recurrence is multiplied by the same factor $\lambda$ at each step.
Normalizing $m_0=1$, we write $m_1=\mathrm{i}\mu$ and $m_2=\frac13+\delta$. Under
$k\to-k$, $m_1$ changes sign while $m_0$ and $m_2$ do not, so
$\lambda=1+\lambda_2k^2+\lambda_4k^4+\cdots$,
$\mu=\mu_1k+\mu_3k^3+\cdots$, and $\delta=\delta_2k^2+\cdots$. Substituting into
the recurrence gives the three real equations
\begin{align}
\lambda&=1+(C_k-1)\left(\tfrac13+\omega_2\delta\right)+S_k\,\omega_1\mu,\\
\lambda\mu&=C_k\,\omega_1\mu-S_k\left(\tfrac13+\omega_2\delta\right),\\
\lambda\left(\tfrac13+\delta\right)&=C_k\left(\tfrac13+\omega_2\delta\right)+S_k\,\omega_1\mu,
\end{align}
which are solved order by order in $k$, using
$C_k=1-\frac{k^2}{2}+\frac{k^4}{24}+\cdots$, $S_k=k-\frac{k^3}{6}+\cdots$,
$1-\omega_1=1/\tau$, and $1-\omega_2=1/\tau_{\nh}$:
\begin{enumerate}
\item At order $k$, the second equation gives $\mu_1=-\tau/3$, that is,
$m_1=-\mathrm{i}k\tau m_0/3$. This is the first-order Chapman--Enskog stress,
$\Pi_{xy}=-\tau c_s^2\rho_0\partial_x\delta u_y$, Eq.~\eqref{eq:pi1}.
\item At order $k^2$, the first equation gives
$\lambda_2=-\frac16+\omega_1\mu_1=-\frac a3$, the viscous decay with
$\nu=a/3$, independent of $\tau_{\nh}$. The term $-\frac16$ comes from the
second-order streaming term $(C_k-1)/3$; combined with the post-collision stress,
$\omega_1\mu_1=-(\tau-1)/3$, it gives $\nu=(\tau-\frac12)/3$, the shift $-\Delta t/2$
of Eq.~\eqref{eq:viscosity}.
\item At order $k^2$, the third equation gives
$\delta_2=\frac23\tau_{\nh}\lambda_2=-\frac{2a\tau_{\nh}}{9}$. The third moment lags
behind its equilibrium, $m_0/3$, which decays with the mode; the lag is
proportional to $\tau_{\nh}$. This is the first appearance of $\tau_{\nh}$.
\item At order $k^3$, the second equation gives
\begin{equation*}
\mu_3=\tau\left[-\lambda_2\mu_1-\frac{\omega_1\mu_1}{2}+\frac1{18}-\omega_2\delta_2\right]
=-\frac{\tau\left(4a^2-8ab+1\right)}{36}:
\end{equation*}
the lag of the third moment corrects the stress at third order in $k$.
\item At order $k^4$, the first equation gives
\begin{equation*}
\lambda_4=\frac1{72}-\frac{\omega_2\delta_2}{2}+\omega_1\mu_3-\frac{\omega_1\mu_1}{6}
=-\frac{a\left(4a^2-8ab-2a+1\right)}{36}.
\end{equation*}
\end{enumerate}
Since $\log\lambda=\lambda_2k^2+\left(\lambda_4-\lambda_2^2/2\right)k^4+\Order(k^6)$,
\begin{equation}
\log\lambda(k)
=-\frac{a}{3}k^2
+\frac{a}{9}\left[a(2b-a)-\frac14\right]k^4
+\Order(k^6).
\label{eq:shear_eigenvalue}
\end{equation}
Odd powers of $k$ are absent because $\lambda$ is even in $k$.

\paragraph{Modified equation}
A Fourier mode of
$\partial_tu_y=\nu\partial_x^2u_y+\kappa_4\partial_x^4u_y+\cdots$ is multiplied in one
time step by $\exp(-\nu k^2+\kappa_4k^4+\cdots)$. Identifying this factor with the
hydrodynamic eigenvalue, Eq.~\eqref{eq:shear_eigenvalue} yields
\begin{equation}
\partial_tu_y
=\frac{a}{3}\partial_x^2u_y
+\frac{a}{9}\left[a(2b-a)-\frac14\right]
\partial_x^4u_y
+\Order(\partial_x^6).
\label{eq:shear_modified_appendix}
\end{equation}
This is the equivalent equation that governs the transverse momentum once the
kinetic modes have decayed, as in the Chapman--Enskog analysis; at order $k^2$
the two analyses give the same viscosity. Unlike that analysis, the present
calculation is restricted to linear perturbations about rest and therefore
does not contain nonlinear terms such as $E^{\mathrm{GI}}_\alpha$. Thus,
$\nu=a/3$ is independent of $\tau_{\nh}$, whereas the hyperviscous coefficient
is an affine function of it, with slope $2a^2/9$; BGK ($b=a$) and RLBM ($b=1/2$) are
two points of this line. Negative values of the coefficient multiplying
$\partial_x^4u_y$ add small-scale damping; positive values reduce that damping
and may promote finite-resolution instability.

\paragraph{Wave vector along a lattice diagonal}
For D2Q9 and $\mathbf K=(q,q)$, with $K^2=2q^2$, the transverse velocity is along
$(-1,1)$. Streaming multiplies population $i$ by
$\exp[-\mathrm{i}q(c_{ix}+c_{iy})]$: the factor is $e^{-\mathrm{i}q}$ for $(1,0)$ and
$(0,1)$, $e^{\mathrm{i}q}$ for $(-1,0)$ and $(0,-1)$, and $1$ for the
populations $(-1,1)$ and $(1,-1)$, which move perpendicular to $\mathbf K$. The
transverse mode is odd under the reflection $x\leftrightarrow y$, which leaves
$\mathbf K$ unchanged. The odd combinations
\begin{equation}
h_1=\delta f_{(0,1)}-\delta f_{(1,0)},\quad
h_2=\delta f_{(0,-1)}-\delta f_{(-1,0)},\quad
h_3=\delta f_{(-1,1)}-\delta f_{(1,-1)}
\end{equation}
are equivalent to the three moments
\begin{equation}
\begin{aligned}
j&=\sum_i\left(c_{iy}-c_{ix}\right)\delta f_i=h_1-h_2+2h_3,\\
\sigma&=\sum_i\left(c_{ix}^2-c_{iy}^2\right)\delta f_i=-(h_1+h_2),\\
r&=-\frac12\sum_i c_{ix}c_{iy}\left(c_{iy}-c_{ix}\right)\delta f_i=h_3,
\end{aligned}
\label{eq:diag_moments}
\end{equation}
where $j=\sqrt2\,\rho_0u_\eta$ is proportional to the transverse momentum,
$\sigma=-2\Pi_{\xi\eta}$ is proportional to the viscous stress in the coordinates
$(\xi,\eta)$ aligned with and normal to $\mathbf K$, and
$r=\frac12(Q_{xxy}-Q_{xyy})$ is a third-order moment carried only by the
diagonal velocities. The linearized equilibrium gives $\sigma^{\ee}=0$ and, by
Eq.~\eqref{eq:lin_q_eq}, $r^{\ee}=\frac12\rho_0c_s^2(\delta u_y-\delta u_x)=j/6$.
Equation~\eqref{eq:lin_moment_relax} gives the collision step
\begin{equation}
j^*=j,\qquad
\sigma^*=\omega_1\sigma,\qquad
r^*=\frac j6+\omega_2\left(r-\frac j6\right),
\end{equation}
and streaming, $h_1'=e^{-\mathrm{i}q}h_1^*$, $h_2'=e^{\mathrm{i}q}h_2^*$, $h_3'=h_3^*$,
gives, with $C_q=\cos q$ and $S_q=\sin q$,
\begin{align}
j'&=C_qj^*+\mathrm{i}S_q\sigma^*+2(1-C_q)r^*,\\
\sigma'&=C_q\sigma^*+\mathrm{i}S_qj^*-2\mathrm{i}S_qr^*,\\
r'&=r^*.
\end{align}
The recurrence is again closed. With $j=1$, $\sigma=\mathrm{i}\mu$, and
$r=\frac16+\varrho$, the mode equations are
\begin{align}
\lambda&=C_q-S_q\,\omega_1\mu+2(1-C_q)\left(\tfrac16+\omega_2\varrho\right),\\
\lambda\mu&=C_q\,\omega_1\mu+S_q\left(\tfrac23-2\omega_2\varrho\right),\\
\lambda\left(\tfrac16+\varrho\right)&=\tfrac16+\omega_2\varrho,
\end{align}
and the same order-by-order solution gives $\mu_1=2\tau/3$,
$\lambda_2=-2a/3$, $\varrho_2=a\tau_{\nh}/9$, and
$\lambda_4=\frac{2a^2}{9}(1-2a+b)$, in powers of $q$. With $q^2=K^2/2$, the
viscosity is again $a/3$, so it is isotropic, and
\begin{equation}
\log\lambda=-\frac a3K^2+\frac{a^2}{18}\left(b-2a\right)K^4+\Order(K^6).
\label{eq:shear_eigenvalue_diag}
\end{equation}
The term $-a/36$ of Eq.~\eqref{eq:shear_eigenvalue} cancels in this
direction, and the diagonal hyperviscous coefficient is only of order $a^2$.

\paragraph{Arbitrary direction}
For an arbitrary direction, the $K^4$ coefficient is a quartic form in
$(K_x,K_y)$ invariant under the symmetry group of the square lattice, hence a
combination of $K_x^4+K_y^4$ and $K_x^2K_y^2$. With
$\mathbf K=K(\cos\theta,\sin\theta)$, $\cos^4\theta+\sin^4\theta=(3+\cos4\theta)/4$ and
$\cos^2\theta\sin^2\theta=(1-\cos4\theta)/8$, so the coefficient takes the form
$\bar\kappa_4+\Delta\kappa_4\cos4\theta$. The two constants follow from
Eqs.~\eqref{eq:shear_eigenvalue} and \eqref{eq:shear_eigenvalue_diag}, giving
Eq.~\eqref{eq:kappa_theta}. We confirmed this form at intermediate angles from
the eigenvalues of the full linearized $9\times9$ D2Q9 operator $M(\mathbf K)$, and
the axis and diagonal coefficients independently by expanding the characteristic
polynomials of the two $3\times3$ recurrences; both scripts are in the
repository listed under Data availability. The diagonal result is specific to
D2Q9; on D3Q19 and D3Q27 the reduced dynamics of oblique modes need not coincide
with it, and fully three-dimensional higher-order errors may retain other
lattice-dependent anisotropy.

\end{appendices}

\bibliographystyle{elsarticle-num}
\bibliography{references}

\end{document}